\documentclass[12pt]{article}
\usepackage[left=1in,top=1in,right=1in,nohead]{geometry}
\usepackage{enumerate}
\usepackage{graphicx}
\usepackage{subfig}
\usepackage{amsmath}
\usepackage{amssymb}
\usepackage{mathrsfs}
\usepackage{multirow}
\usepackage{bm}
\usepackage{color}
\usepackage{jheppub}
\usepackage{verbatim}
\usepackage{placeins}

\newcommand{\be}{\begin{equation}}
\newcommand{\ee}{\end{equation}}
\newcommand{\bea}{\begin{eqnarray}}
\newcommand{\eea}{\end{eqnarray}}
\newcommand{\bra}[1]{| #1 \rangle}

\def\gsim{\, \rlap{$>$}{\lower 1.1ex\hbox{$\sim$}}\,}

\begin{document}

\title{Violations of the Born rule in cool state-dependent horizons}

\author[a]{Donald Marolf}

\author[b]{and Joseph Polchinski}

\affiliation[a]{Department of Physics, University of California,\\ Santa Barbara, CA 93106-9530 USA.}

\affiliation[b]{Kavli Institute for Theoretical Physics, University of California,\\
 Santa Barbara, CA 93106-4030 USA.}
%
%% e-mail addresses:
%
%% e-mail addresses:
\emailAdd{marolf@physics.ucsb.edu}

\emailAdd{joep@kitp.ucsb.edu}

\abstract{The black hole information problem has motivated many proposals for new physics.  One idea, known as state-dependence, is that quantum mechanics must be generalized to describe the physics of black holes, and that fixed linear operators do not provide the fundamental description of experiences for infalling observers.  Instead, such experiences are to be described by operators with an extra dependence on the global quantum state.  We show that any implementation of this idea strong enough to remove firewalls from generic states
% added: from generic states
requires massive violations of the Born rule. We also demonstrate a sense in which such violations are visible to infalling observers involved in preparing the initial state of the black hole.  We emphasize the generality of our results; no details of any specific proposal for state-dependence are required.}

\maketitle

%%%%%%%%%%%%%%%%%%%%%%%%%%%%%%%%%%%%%%%%
%
\section{Introduction}
%
%%%%%%%%%%%%%%%%%%%%%%%%%%%%%%%%%%%%%%%%
\label{sec:intro}

\baselineskip=17pt

The black hole information problem is a fundamental tension between the presumed finite density of black hole states and the presumed validity of effective field theory for infalling observers.  The issue was recently re-emphasized in the arguments of~\cite{Almheiri:2012rt,Almheiri:2013hfa,Marolf:2013dba,Bousso:2013wia}, which showed that black hole complementarity \cite{Susskind:1993if} alone is insufficient to resolve the tension and which also critiqued a variety of other approaches.  Ref.~\cite{Almheiri:2012rt} argued that three widely-held assumptions are mutually inconsistent: (1) the existence of an S-matrix describing black hole formation and evaporation, (2) the validity of effective field theory outside the black hole, and (3) the absence of drama for an infalling observer.  We emphasize that  a contradiction results even when requirement (2) is used only to order one accuacy for simple quantities.

The form of drama envisioned in~\cite{Almheiri:2012rt} was what we might call `fiery drama,' a wall of high energy particles~\cite{Braunstein:2009my}, though we will include in same category any strong modification of the smooth horizon geometry, including braney drama (\cite{Mathur:2012zp},  but without fuzzball complementarity \cite{Mathur:2012jk,Mathur:2013gua}), stringy drama~\cite{Giveon:2012kp,Giveon:2013ica,Giveon:2015cma} and \cite{Silverstein:2014yza,Dodelson:2015toa}, or other scenarios~\cite{Giddings:1992hh,Chapline:2000en,Mazur:2001fv,Winterberg,Davidson:2011eu}.
Many attempts to evade this conclusion have instead substituted what we might call `quantum drama,' where the rules of quantum mechanics are modified or augmented for the infalling observer.  Ideas in this class include the black hole final state proposal~\cite{Horowitz:2003he,Lloyd:2013bza}, limits on quantum computation combined with strong complementarity~\cite{Harlow:2013tf}, ER = EPR~\cite{Maldacena:2013xja}, and state-dependent observables as developed in~\cite{Papadodimas:2012aq,Papadodimas:2013wnh,Papadodimas:2013jku,Papadodimas:2013kwa,Papadodimas:2015xma,Papadodimas:2015jra} (related ideas appeared in~\cite{Nomura:2012sw,Nomura:2013gna,Nomura:2012ex} and~\cite{Verlinde:2012cy,Verlinde:2013uja,Verlinde:2013vja,Verlinde:2013qya}).

In this note we clarify the way in which the last of these ideas, state-dependence, modifies ordinary quantum mechanics.  Ref.~\cite{Harlow:2014yoa} has already critiqued the proposal of~\cite{Papadodimas:2012aq,Papadodimas:2013wnh,Papadodimas:2013jku,Papadodimas:2013kwa,Papadodimas:2015xma,Papadodimas:2015jra} in some detail.  Our point of view is largely the same, but we wish to step back from the details of specific proposals.  We will argue that {\it any} framework relying on state-dependence alone to eliminate firewalls in generic states implies large violations of the Born rule, and that these can be visible to infalling observers.  In the remainder of this introduction we review the idea of state-dependence.  In \S2 we demonstrate Born rule violation, and we close with some final discussion in \S3.  To avoid infra-red issues and for comparison with \cite{Papadodimas:2012aq,Papadodimas:2013wnh,Papadodimas:2013jku,Papadodimas:2013kwa,Papadodimas:2015xma,Papadodimas:2015jra} we focus on asymptotically anti-de Sitter (AdS) black holes below. Some calculations and refinements are set aside in the Appendices.

While textbook quantum mechanics associates any physical observable with some fixed linear operator on the Hilbert space, state-dependence would change this for observers falling into black holes.  To see how state-dependence enters the discussion, suppose that one postulates i) that any near-equilibrium state of the black hole has a smooth horizon as predicted by effective field theory and ii) that such states are typical in the Haar measure associated with the unitary group on the finite-dimensional Hilbert space of fixed energy.  Refs.~\cite{Bousso:2013wia,Marolf:2013dba} have shown that one can find a basis of states (eigenstates of number operators of the Hawking modes) such that almost all have firewalls -- high energy excitations of many modes as seen by an infalling observer.  If there were a linear projection operator onto states with such excitations, it would then be very close to the identity.  So typical states would have firewalls in contradiction to the assumption.  State-dependence avoids this by allowing the projection defining the excitation to vary in a nonlinear way as one moves across the Hilbert space.  The physical interpretations of the original basis states can then be unrelated those of other states.

The term `state-dependence' describes in a precise way the nature of the observables being considered, but its innocuous sound hides the radical nature of the idea.  It is easily conflated by the unwary with the more general and usually benign property of background-dependence.  A simple example of the latter is given by the collective coordinate quantization of solitons~\cite{Goldstone:1974gf,Gervais:1975pa,Callan:1975yy}.  Here there are many classical background solutions, related by translation.  Internal excitations of the soliton are defined relative to the center of mass, and so are background-dependent.  However, they remain linear operators in the Hilbert space.  The change of variables from the original path integral fields to the center-of-mass variables is nonlinear in the {\it fields}, but fully linear in terms of the Hilbert space structure.  Thus one may assemble the naturally-defined background dependent operators into a background-independent operator.  In contrast, as shown above, using state-dependence to eliminate firewalls requires that such an assembly into a linear operator be impossible even in principle.

Other familiar examples of background dependence are similarly field-dependent, not state-dependent~\cite{Harlow:2014yoa}.  For example, measurements that are conditioned on earlier measurements remain linear --- see \S5 of~\cite{Harlow:2014yoa}, and \S3.1 of~\cite{Preskill}.  We know of no situation in ordinary quantum mechanics where observables are fundamentally state-dependent.

Ref.~\cite{Papadodimas:2015jra} has suggested that state-dependence might be needed for the construction of bulk fields even exterior to a black hole in terms of a dual CFT.  However, Ref.~\cite{Heemskerk:2012mn} argued that this construction is simply field-dependent, with the background-dependence arising from the nonlinearity in the fields of the bulk-boundary map.  The construction~\cite{Kabat:2012av,Heemskerk:2012mn,Kabat:2013wga,Kabat:2015swa} was given as an expansion in powers of fields, whose convergence might appear to be an issue.  But it is better to think of this as a differential construction, allowing one to move continuously from one background to another until one encounters a natural barrier, in the spirit of analytic continuation.  One known natural barrier is a black hole horizon.   We know of no evidence for any other barrier, nor any reason to believe that if one did exist it could be surmounted by state-dependence.

Finally, we should note that if state-dependence does turn out to be a necessary property of quantum gravity -- a feature and not a bug -- it will be essential to make clear both its precise nature and the obstacles to its consistent implementation.  The discussion below would then provide a step in this direction.

\section{State-dependence and the Born rule}
\label{moreSD}

We now argue that any state-dependence strong enough to allow the horizons of typical black holes to be experienced as vacuum by infallers leads to gross violations of the Born rule.  Furthermore, there is a sense in which these violations can be visible to sufficiently powerful infalling observers.  We assume only 1) that simple experiences of infallers in typical pure quantum states with sufficiently large fixed energy $E_0$ are, with high probability, governed by effective field theory in the Hartle-Hawking state $|{\rm HH} \rangle$ on the classical black hole background, (2) that operators at the AdS boundary obey the usual rules of quantum mechanics (e.g., because they can be mapped directly to a dual CFT); in particular, they --- or at least bounded functions thereof --- are linear operators defined globally on the entire Hilbert space, (3) that the Bekenstein-Hawking entropy $S_{\rm BH}$ describes the logarithmic density of states at energies $E_0$ for which (1) holds. The detailed constructions of \cite{Papadodimas:2012aq,Papadodimas:2013wnh,Papadodimas:2013jku,Papadodimas:2013kwa} or \cite{Verlinde:2012cy,Verlinde:2013uja,Verlinde:2013vja,Verlinde:2013qya} are not required.  Indeed, the argument below applies to any proposal that achieves the above goals whether or not state-dependence is explicitly involved.

Following the terminology of \cite{Papadodimas:2012aq,Papadodimas:2013wnh,Papadodimas:2013jku,Papadodimas:2013kwa}, we refer to the states in (1) as being in near-equilibrium.  Just how near is quantified by the probability $\epsilon$ of an infaller experiencing an excitation relative to $|{\rm HH}\rangle$.  The term `typical' is similarly quantified by saying that these states form a set of measure $1-\tilde \epsilon$ with respect to the normalized round measure on the unit-sphere of normalized states with energy $E < E_0$.  Since we consider a theory of gravity, this energy can be measured at the boundary and so by (2) defines a state-independent operator.   It is implicit in (1) that $\epsilon, \tilde \epsilon$ vanish in the limit where the AdS scale $\ell_{AdS}$ is large compared with the Planck scale $\ell_{P}$.

We note that (1) includes the results of simple manipulations of these states by infallers.  If these manipulations are performed outside the black hole, they may be expressed in terms of operators that, again by (1) may be mapped to operators on the AdS boundary using the techniques of \cite{Banks:1998dd,Balasubramanian:1998de,Hamilton:2006az,Kabat:2011rz,Kabat:2012av,Kabat:2012hp,Heemskerk:2012mn,Kabat:2013wga,Kabat:2015swa,Morrison:2014jha}.
Using (2) then shows that simple effective field theory operators outside the horizon -- which might for example add extra particles or correlated small sets thereof to $|{\rm HH}\rangle$ -- correspond to state-independent operators whose action on near-equilibrium states differs only be terms of order $\epsilon.$

For concreteness, in all cases below we work in asymptotically AdS$_d$ spacetimes for $d \ge 3$ and assume the existence of a dual conformal field theory with order $N^2 \sim \left(\frac{\ell_{AdS}}{\ell_P}\right)^{k}$ fields for some positive $k$.
%\DM{[Did I get that right?]}
We consider states near some energy $E_0$ above the Hawking-Page transition,  where the density of states is dominated by large global AdS-Schwarzschild (or BTZ) black holes (in Appendix A we will have use for  black holes that are smaller but still stable).  As usual, we take their area-radius to be $r_0$ and their temperature to be $T_0$.  We will use ${\cal H}_{E_0}$ to refer to the space of pure states with energy $E < E_0$. Taking $G_{\rm N}$ to be the bulk Newton constant, to leading order in $1/G_{\rm N}$ the density of states is $e^{S} = e^{A/4G_{\rm N}}$ in terms of the horizon area $A$.

Some arguments in~\cite{Almheiri:2012rt,Almheiri:2013hfa,Marolf:2013dba,Bousso:2013wia} applied to black holes that are maximally entangled with another system, and others to pure states.  In standard quantum mechanics these settings are equivalent due to the fact that observables are defined by fixed linear operators. Properties of entangled black holes are thus determined by the reduced density matrix obtained by tracing out the system with which it is entangled.  Since we now consider relaxing the rules of quantum mechanics, these
need to be considered separately.  In the present section we consider pure states, and in conclusions we discuss some issues of entanglement.

\subsection{Violations of the Born rule}
\label{sec:Born}

The Born rule of quantum mechanics states that the probability to observe a particular experimental outcome $\chi$ in a quantum state $|\psi \rangle $ is $\sum_i |\langle \chi,i | \psi \rangle|^2$, where $|\chi,i\rangle$ are a complete set of orthonormal states in which $\chi$ occurs with probability one.  In particular, given two physically exclusive experimental outcomes $\chi, \chi'$ that occur with probability one in states $|\chi \rangle$, $|\chi' \rangle$, we must have $\langle \chi |\chi' \rangle =0$.  We show this condition to be strongly violated by any proposal satisfying the assumptions above.

The presence of some violation of the Born rule follows from the definition of state-dependence.  If the Born rule held exactly we could construct a linear projector $P =
\sum_i | \chi, i \rangle \langle \chi,i|$ onto the outcome $\chi$, and this is excluded if typical states are firewall-free.  On the other hand, the firewall basis states defined above have only $O(e^{-S/2})$ overlap with typical states since the latter are linear combinations of $O(e^S)$ basis states.  But by applying an argument from \S5 of~\cite{Almheiri:2013hfa} we demonstrate much larger violations of the Born rule below.

Before proceeding with a precise discussion, we discuss a toy example so as to identify the issues.  Consider $b_\omega$, a fermionic mode of definite frequency in the static geometry external to the black hole. In the infalling vacuum state, this is entangled with an inner mode $\tilde b_\omega$,
\be
\label{eq:psi}
|\psi\rangle =  \frac{ |\tilde 0,0 \rangle + x^{1/2} |\tilde 1, 1\rangle}{(1+x)^{1/2}} \,,\quad x = e^{-\omega/T_0} \,.
\ee
Now let $U = \exp(i\pi b^\dagger_\omega b_\omega)$, where we take discrete normalization for convenience.  Then
\be
U |\psi\rangle =  \frac{ |\tilde 0,0 \rangle - x^{1/2} |\tilde 1, 1\rangle}{(1+x)^{1/2}} \,.
\ee
Computing the inner product $\langle \psi | U | \psi\rangle$, the state $ U | \psi\rangle$ has probability $\left[(1-x)/(1+x)\right]^2$ to be in its ground state $|\psi\rangle$ and so probability $4x/(1+x)^2$ to be excited.  We see that $U$ maps the space of typical states to a space of states with nonvanishing excitation probability.  However, the state $U| \psi\rangle$ would seem to be every bit as typical as $|\psi\rangle$, contradicting the assumption that typical states have vacuum near the horizon.

The problem that we have found is not precisely as was advertised above.  One difference is that $| \psi\rangle$ and $ U | \psi\rangle$ are not orthogonal.  But this can be accomplished by a similar construction using two modes, $b_\omega, b'_\omega$ with identical frequencies, both in the same state \eqref{eq:psi}.  For
\be
U = \exp(i\theta b^\dagger_\omega b_\omega -  i\theta b'^\dagger_\omega b'_\omega - i\pi b^\dagger_\omega b_\omega  b'^\dagger_\omega b'_\omega) \,, \label{unitary}
\ee
one finds
\be
\label{eq:2modes}
\langle \psi, \psi | U | \psi, \psi\rangle = \frac{1 + 2x \cos\theta - x^2}{(1+x)^2} \,.
\ee
For $x$ not too small there is a choice of $\theta$ where this vanishes.   For two modes with distinct frequencies $\omega, \omega'$ both in \eqref{eq:psi} (with distinct $x,x'$) one can similarly obtain an orthogonal state by separately tuning the phases in \eqref{unitary} separately.

We now have a single state vector with two orthogonal physical interpretations.  If one declares that all such states are vacuum, one encounters the frozen vacuum problem~\cite{Bousso:2013ifa}.  There are operations by the infalling observer that should produce excitations in effective quantum field theory, but they do not, and so the rules of physics are changed in an observable way at the horizon.

However, this toy example has an important limitation~\cite{Papadodimas:2013jku}.  Modes of definite frequency are distributed over all space and time.  As a result, they cannot be measured by the infalling observer, and in particular the interior part $\tilde b_\omega$ cannot.  In order for a phenomenon to be seen by this observer, we must restrict to modes with limited support in space and time.  These have a nonzero width in frequency, and so their number operator does not commute with the Hamiltonian.\footnote{This is in part a good thing, because for a chaotic system with finite density of states one expects the only operators commuting with the Hamiltonian to be functions of the global conserved charges.  This is a sign that effective field theory breaks down when used to describe operators which act entirely within a subspace of energies with width of order $e^{-S_{\rm BH}}$. Below, we use operators whose energy uncertainties are small, but not exponentially small. Their description by effective field theory is self-consistent.}  Then $U| \psi\rangle$ does not have the same energy as $| \psi\rangle$
(and generically its energy is greater).  Our goal will be to show that the increase in energy is small enough that most typical states are nearly parallel to states having an orthogonal physical interpretation.  This gives a large and observable violation of the Born rule.

Consider now a normalizeable wave-packet $b$ supported outside the horizon. We also use the symbol $b$ to denote the operator obtained by taking the symplectic product of our wave packet (i.e., the Klein-Gordon product in the case of a scalar field) with the associated linearized quantum field, with conventions set so that $b$ would be the usual annihilation operator for an eigenmode of positive frequency.  Using the bulk-boundary dictionary, we can write $b$ in terms of an operator at the AdS boundary~\cite{Banks:1998dd,Balasubramanian:1998de,Hamilton:2006az,Kabat:2011rz,Kabat:2012av,Kabat:2012hp,Heemskerk:2012mn,Kabat:2013wga,Kabat:2015swa,Morrison:2014jha}.\footnote{For large angular momenta, the construction becomes complicated due to large grey-body factors, but we will not need such modes.}  To be precise, this construction applies in typical states, for which the external geometry is an AdS-Schwarzschild black hole; let the definition be extended to atypical states in an arbitrary manner consistent with linearity and preserving the operator norm of $b$.  For a fermionic field, $b$ is bounded and thus a globally-defined linear operator on the Hlibert space.

We are interested in smooth wave-packets of small width $\Delta \omega$ in frequency space near some $\omega_0 > 0$ of order $T_0$ and which are supported in a Rindler-like region near the black hole horizon.  In flat space, such modes can be localized in a region of space of size $\Delta x \sim 1 /\Delta \omega$, in which the mode executes $n \sim \omega_0/\Delta \omega$ oscillations.  A similar principle applies in the black hole background, as may be seen by considering an asymptotically flat black hole that emits such a mode into the surrounding flat-space region.  Running the evolution backward to a slice of Killing time where the mode is confined to the near-horizon region leaves $\Delta \omega$, $\omega_0$, and the number of oscillations unchanged.  So we have $\Delta \omega \sim \omega_0/n \sim T_0/n$ in this region as well (whether the desired black hole is asymptotically flat or asymptotically AdS).  What changes, however, is that modes of definite Rindler frequency oscillate logarithmically in the near-horizon region, executing an infinite number of oscillations outside the horizon.  We restrict our modes to those supported at more than a Planck length $\ell_{\rm P}$ of proper distance from the bifurcation surface. The fact that the Rindler region for a large black hole extends to proper distances of order $\ell_{\rm AdS}$ then bounds the maximum number of oscillations, but still allows $n \sim \ln(\ell_{\rm AdS}/\ell_{\rm P}) \sim \ln N$.

We will make use of a set of outside modes $b_i$ for $i = 1 \dots N_{\rm modes}$ associated with a fermionic semi-classical quantum field. We choose $N_{\rm modes}$ to be small enough that we may take all modes to be peaked near the same frequency $\omega_0 \sim 1/r_0 \sim T_0$, and to differ only in their radial profile so that all $b_i$ are naturally intercepted by a common infalling observer.

We construct unitary transformations by the same strategy as above.  Taking the unitary to be diagonal in the occupation number basis to minimize the commutator with the Hamiltonian, there are $2^{N_{\rm modes}} - 1$ phases to choose.  If we wish to prepare $m$ mutually orthogonal states $U_I | \psi\rangle$, there are $m(m-1)$ conditions on these phases.  For linear equations, the largest possible number would be $m \sim 2^{N_{\rm modes}/2}$ states.  Some numerical experimentation indicates that $\ln m \propto N_{\rm modes}$ for the present problem as well.\footnote{The analysis is not completely trivial because the state does not factorize for modes of nonzero $\Delta \omega$, but it is a small perturbation of this. }

For a given unitary $U_I$ constructed as above we wish to study the states
$ U_I {\cal H}_{E_0}$ obtained by applying $U_I$ to all states in ${\cal H}_{E_0}$.  By construction infalling vacuum states in ${\cal H}_{E_0}$, which are Haar-typical, map to excited states under $U_I$ for each $I$.
Recall that the unitary $U_I$ is a state-independent operator since it acts only on the external fields.
The action of the unitary raises the mean energy of typical states by order
\be
\Delta E_{\rm mean} \sim \frac{N_{\rm modes} (\Delta\omega)^2 }{T_0} \sim \frac{N_{\rm modes} T_0}{n^2}  \label{deltae} \,.
\ee
For appropriate choices of modes the probability of a fluctuation $\Delta E$ above the mean is
\be
e^{- O( n^2 \Delta E^2 /N_{\rm modes} T_0^2)} \,. \label{var}
\ee
Similar calculations are given in a slightly different context in Appendix A.  The width $\Delta E_{\rm var} \sim N_{\rm modes}^{1/2}T_0 / n$ in (\ref{var}) is larger than the shift in the mean in~(\ref{deltae}) so we focus  $\Delta E_{\rm var}$.
If we project $U_I {\cal H}_{E_0}$ onto ${\cal H}_{E_+}$ for $E_+ = E_0 + \lambda \Delta E_{\rm var}$ with $\lambda$ a constant somewhat larger than one, this projection $P$ will leave a typical state invariant up to corrections that vanish as $e^{-O(\lambda^2)}$.

The space $P  U_I  {\cal H}_{E_0}$ is a subspace of ${\cal H}_{E_+}$.  The respective dimensionalities are
\be
\dim (P  U_I  {\cal H}_{E_0}) = \dim ({\cal H}_{E_0}) = e^{S_0} \, , \quad \dim ({\cal H}_{E_+} )= e^{S_0 + \lambda \Delta E/T_0}  \,.
\ee
In the first equality we use the fact that $P$ has negligibly small kernel for generic $U$, as is readily seen by counting equations and unknowns, but any kernel associated with non-generic $U$ would shrink as we increase $\lambda$ and  a small kernel would not affect the result.
The {\it difference} of dimensions is enormous, as noted in~\cite{Papadodimas:2015jra}, but the {\it ratio} is close to unity provided
\be
\delta \equiv \frac{\lambda \Delta E}{T_0} = \frac{\lambda N_{\rm modes}}{n^2} \ll 1 \,. \label{delta}
\ee
We will see that this is what leads to Born rule violations.  Thus we obtain parametrically large violations by taking $n$ large with fixed $\lambda$ and $N_{\rm modes}$.  In the present construction, we have noted that $n \sim \ln N$, but in Appendix A we give another construction where $n$ is a power of $N$.

The key observation is that a typical state $|\psi_+ \rangle$ in ${\cal H}_{E_+} $ can be written
\be
|\psi_+ \rangle = \cos \theta |\psi_1 \rangle + \sin\theta |\psi_2 \rangle, \quad {\rm with} \quad \cos^2\theta \cong \frac{\dim (P  U_I  {\cal H}_{E_0}) }{\dim ({\cal H}_{E_+} )} = e^{-\delta}\,, \label{angle}
\ee
where $ |\psi_1 \rangle$ and $|\psi_2 \rangle$ are normalized states  in $P  U_I  {\cal H}_{E_0}$ and its orthogonal complement ${\cal H}_\perp = {\cal H}_{E_+}/P  U_I  {\cal H}_{E_0}$ respectively.  This is readily seen by expanding in a basis, and the variations of $\theta$ are actually quite small due to the high dimensionality.  So for small $\delta$ almost all states in ${\cal H}_{E_+} $ are nearly parallel to states in $P  U_I  {\cal H}_{E_0}$.

We are essentially done.  We have just seen that almost all states in ${\cal H}_{E_+}$ are nearly parallel to states in $P  U_I  {\cal H}_{E_0}$.  And almost all states in $P  U_I  {\cal H}_{E_0}$ are nearly parallel to states in $ U_I  {\cal H}_{E_0}$, the exceptions arising from states in $ U_I  {\cal H}_{E_0}$ whose energy happens to vary upward in energy far enough that the projection $P$ has a large effect.  Finally, almost all states in $ U_I  {\cal H}_{E_0}$ are excited, the exception being those obtained from atypical states in ${\cal H}_{E_0}$.  So almost all states in ${\cal H}_{E_+}$ are assumed to be in vacuum, but almost all are nearly parallel to excited states in $ U_I  {\cal H}_{E_0}$: this is the Born rule violation.\footnote{This contradicts an argument in sections 8.3.1 and 8.3.2 of \cite{Papadodimas:2015jra}, which considers a tiny neighborhood of a subspace like $P  U_I {\cal H}_{E_0} \subset {\cal H}_{E_+}$.  The key point is that almost all the volume of a high-dimensional sphere is very close to the equator; the calculation in that work ignored the curvature of the sphere.
}  A precise version of this statement is derived in Appendix B.

\subsection{Observing the violation}
\label{sec:obs}

To describe the sense in which the above violation is visible to infallers, it is again useful to think of our AdS system as a conformal field theory sitting inside a larger laboratory.  In fact, we may imagine this laboratory to contain many copies of the CFT which may be manipulated at will by a sufficiently powerful experimenter\footnote{Despite having states with distinct infaller-outcomes that are extremely close together, we assume the outcomes are sufficiently continuous that an ardent experimenter can produce a state with high probability to give the desired results.}.  After performing some number of tests and manipulations, the experimenter then injects herself into some particular copy of the CFT so as to become a bulk AdS observer and fall into the black hole.  For clarity, we take the target CFT for this injection to be unentangled with anything else in the laboratory. We imagine this setting to be analogous to having an extremely powerful observer in asymptotically flat space, who may test and prepare as many black holes as she likes before finally choosing one into which she will jump.  The advantage of moving the observer completely outside the gravitating spacetime is simply that it frees us from discussing possible practical constraints on the extent to which the black holes (and the ensuing Hawking radiation) might be manipulated.

We assume that our observer knows the full theory describing the bulk.  Her goal is to test this theory, and in particular to produce strong evidence that the experiences of infallers violate the Born rule.  She would like to verify that there is a set of states $\{|I \rangle \}$, labeled by a set of distinct outcomes $I$, which nearly coincide in the CFT Hilbert space but where bulk infalling observers in each $|I \rangle$ have probability essentially one to experience the corresponding $I$.  Verifying that two given CFT states $|I \rangle$ and $|J \rangle$ nearly coincide is straightforward.
She merely entangles the CFT with a spin -- a $j=1/2$ representation of SU(2) -- elsewhere in her laboratory to prepare the state $|I \rangle |{\rm up} \rangle + |J \rangle |{\rm down} \rangle$ and then measures the spin in the basis $|\pm \rangle = |{\rm up} \rangle \pm |{\rm down} \rangle.$  Since $|I \rangle$ and $|J\rangle$ are nearly equal, she finds the state $|{\rm up}\rangle$ with overwhelming probability.  Note that she is free to repeat this experiment as many times as she desires by making use of the many copies of the CFT in her lab, verifying that the probability is very high and showing that the same conclusion holds for every pair of states taken from the set $\{|I \rangle \}$.

On the other hand, almost by definition it is impossible for her to verify that all of the states $\{|I \rangle \}$ lead to the predicted outcome for infallers. Here we assume that infallers cannot communicate their experiences back to anyone outside the black hole.  We also assume that once our observer decides to enter some black hole she will be destroyed in the singularity and thus unable to probe any further black holes. This prohibits her from testing more than one of the states $|I \rangle$.

But what she {\it can} do is to choose a complicated $|I \rangle$, where the binary representation of $I$ encodes a message that requires a large number of bits.    By verifying that the modes are excited as expected, she obtains strong evidence that the algorithm mapping state $|I \rangle$ to outcome $I$ is correct, and thus that the physics of infallers is governed by massive violations of the Born rule.
The strength of this evidence, quantified by the number of bits in the message, can be arbitrarily large at large $N$.

We should note some potential obstacles and improvements.  First, if we use the $L=0$ wave modes only, then a single infalling observer cannot measure their state with perfect fidelity, since this is spread over all directions on the sphere~\cite{Freivogel:2014fqa}.  However, we are free to use a linear combination of different partial waves, and a modest number will produce a mode sufficiently localized to be measured with some accuracy.  Alternatively, we could use an error-correction scheme to encode each bit of the message in multiple $L=0$ modes so that the fidelity to retrieve the message remains large even with imperfect fidelity for each mode.

Second, the number of measurable modes that one can excite is of order $\ln N$ in the above construction, and so enormous values of $N$ are needed in order to obtain high statistical certainty that the Born rule is violated.  This should not be a concern, since we are simply establishing a point of principle.  But it is also possible to do better.  Appendix A describes a construction in which the frequency width can be of order $N^{-2}$, drastically reducing the value of $\delta$.   If we keep to a single partial wave, the number of modes visible behind the horizon before the infalling observer hits the singularity is still only of order $\ln N$.  We could multiply this by using many partial waves, but this would lead to very large gray-body factors.  The necessary operators $b$ in the CFT can still be constructed, but they involve terms of order $e^L$ canceling against each other and so this is theoretically complicated.

A better approach is to use a large number of species.  We can cross $AdS_5$ with a stack of $M$ D1-branes, with $M$ a power\footnote{The largest power allowed is determined by a calculation analogous to \cite{Lawrence:1993sg,Frolov:2002qd,Frolov:2003kd,Brown:2012un}.} of $N$, which will thread the black hole.  This gives $M^2$ species: our observer can excite some and not others behind the horizon, and then verify the Born-rule violating prediction.  This construction has two further advantages as well.  First, the excitations are localized at one point on the horizon and so more easily measured.  Second, one may note that choosing modes with small $\Delta \omega$ forces some part of the excitation to be very close to the horizon, and thus to have high energy in the frame of an infalling observer.  Confining the excitation to a D1-brane crossing the horizon immediately makes clear that this excitation can be detected by our infaller without causing her total annihilation.

\section{Discussion}
\label{sec:disc}

Violations of the Born rule do not necessarily imply an intrinsic inconsistency.  It remains possible that this will prove to be a necessary feature of quantum gravity.  But giving up the basic structure of quantum mechanics means that there are many things to check.  In particular, the theory must ultimately give a definite prediction for what the observer will find in the interior of a black hole, for any given state of the system -- including those that describe  black holes entangled with external systems.  The linearity of quantum mechanics allows definite predictions in this case, but in the state-dependent context new information is required.

Consider for example two states such as we have constructed, $|v\rangle$ and $|e\rangle = U|v'\rangle $, which are nearly equal as state vectors but where the first is infalling-vacuum and the second has an excitation at the horizon.  If we entangle the black hole with another qubit and then jump in, what prediction is to be made for the state
\be
\frac{|v,0\rangle + |e, 1\rangle}{\sqrt 2} \,, \label{qubit}
\ee
where $0,1$ are the states of the external qubit?  It is natural to suppose that the probability is $\frac12$ to find the state $v$ and $\frac12$ to find the state $e$.  But what if we write the same state as
\be
\frac{|e+v,0+1\rangle + |e-v, 0-1\rangle}{2\sqrt 2} \, ?
\ee
Since $e-v$ is very close to zero, the black hole state is essentially $(e+v)/2$.  But to make contact with the prediction of~(\ref{qubit}) one must know to separate this state into the specific parts $e$ and $v$, and if just given this state as a sum there is no indication of how to make this split.  Indeed, in the particular formalism of Ref.~\cite{Papadodimas:2013jku} will find a single unitary $U'$ that takes this total state to a vacuum $v''$, and there is no simple connection between the physical interpretation of this state and that given by~(\ref{qubit}).

Refs.~\cite{Gisin, Polchinski:1990py} did note exotic behavior in one version of nonlinear quantum mechanics~\cite{Weinberg:1989us}, but this did not necessarily lead to inconsistency.  Moreover in the present context where the nonlinear behavior is limited to observations behind the horizon, the consequences may be more limited.  However, those papers probed the consistency of the theory only in a limited way.  States were allowed to evolve nonlinearly between observations, but observations were tied to ordinary linear observables in order to use the familiar interpretation of the wavefunction.  In the case at hand, the nonlinearity of the observables is the key issue.

We have not discussed the ER=EPR idea~\cite{Maldacena:2013xja}, although it shares some features with state-dependence.  In particular, the interior geometry depends on the degree of entanglement, which cannot be measured by a linear operator~\cite{Maldacena:2013xja}. So again there is nonlinearity in the observables.  However, in some versions of this idea it is considered that typical states of sufficient complexity might have firewalls~\cite{Susskind:2014ira}.  This could entail a weaker use of state-dependence than assumed above so that our arguments do not immediately apply. The associated modification of quantum mechanics would then be harder to observe.

\section*{Acknowledgements}

We thank the participants of the Aspen Center for Physics summer program ``Emergent Spacetime from String Theory" for useful discussions.  We have particularly benefitted from conversations with Dan Harlow, Kyriakos Papadodimas, Suvrat Raju, and Erik Verlinde.
DM was supported in part by FQXi under grant FQXi-RFP3-1338, the National Science Foundation under Grant No PHY12-05500, and by funds from the University of California.   JP was supported in part by NSF 
PHY11-25915 (academic year) and PHY13-16748 (summer).  This work was completed at the KITP program ``Quantum Gravity Foundations: UV to IR," supported by NSF PHY11-25915.
\appendix
\section{Other constructions}

To reduce the width of the wavepackets in frequency space, and thereby reduce the small parameter $\delta$, we will consider small but stable black holes in anti-de Sitter space.  For example, in $AdS_5 \times S^5$ one can have a ten-dimensional black hole with Schwarzschild radius $r_0 \gsim N^{-2/17} \ell_{\rm AdS}$~\cite{Horowitz:1999uv}.  The number of modes in a given partial wave external to the black hole with frequencies of order $1/r_0$ is $N_{\rm ext} \sim N^{2/17}$, and we can take a basis $b_i$ with frequency widths of order $1/N_{\rm ext} r_0$, or $n \sim N_{\rm ext}$ in the notation of (\ref{deltae}).  The point is that, in order to construct our unitary, we need all of the modes to be external to the black hole on a common time slice.

We wish to determine the effect of the unitary
\be
U = e^{i X} \,,\quad  X = \sum_i \theta_i b_i^\dagger b_i
\ee
on the energy
\be
\langle \psi | e^{-iX} H e^{iX}|\psi \rangle \,.
\ee
Writing
\be
b_i = \int \frac{d\omega}{2\pi} g_i(\omega) b_\omega \,,
\ee
the first two terms in the expansion of the exponential involve
\bea
i[H, X] &=& i\sum_i \theta_i \int \frac{d^2\omega}{(2\pi)^2}  (\omega_1 - \omega_2) g_i^*(\omega_1)g_i(\omega_2) b_{\omega_1}^\dagger b_{\omega_2} \,,
\nonumber\\
-[[H, X],X] &=& -\sum_{ij} \theta_i \theta_j \int \frac{d^3\omega}{(2\pi)^3}  (\omega_1 - \omega_2) g_i^*(\omega_1)g_i(\omega_2) \nonumber\\
&&\qquad\qquad
\left[  g_j^*(\omega_2)g_j(\omega_3) b_{\omega_1}^\dagger b_{\omega_3} - g_j^*(\omega_3)g_j(\omega_1) b_{\omega_3}^\dagger b_{\omega_2} \right]\,.
\eea
Inserting the thermal average
\be
\langle \psi | b_{\omega_1}^\dagger b_{\omega_2} |\psi \rangle = 2\pi \delta(\omega_1 - \omega_2)N(\omega_1)\,,
\ee
the first term vanishes and the second becomes
\be
-\sum_{ij} \theta_i \theta_j \int \frac{d^2\omega}{(2\pi)^2}  (\omega_1 - \omega_2) (N(\omega_1) - N(\omega_2))  g_i^*(\omega_1)g_i(\omega_2)
 g_j^*(\omega_2)g_j(\omega_1) \,.
\ee
Because the packets are narrow, the product  $g_i(\omega) g_j(\omega)$ falls rapidly for $i$ very different from $j$, so we can estimate the sum by setting $i = j$.  Noting that the total area of $
g_i^*(\omega_1)g_i(\omega_2) $ is 1, and that $(N(\omega_1) - N(\omega_2)) \sim (\omega_2 - \omega_1)/T_0$, this is of order
\be
\frac{1}{T_0} \sum_i \theta_i^2  \int \frac{d^2\omega}{(2\pi)^2}  (\omega_1 - \omega_2)^2 |g_i(\omega_1)|^2 |g_i(\omega_2) |^2 \sim \frac{T_0}{N_{\rm ext}^2} \sum_i\theta_i^2 \,.
\ee
For all $\theta_i$ of order one, this is of order
\be
T_0/N_{\rm ext} \,, \label{shift}
\ee
suppressed by a power of $N$ as compared to the $\ln N$ obtained in the construction in the main text.  Higher terms are of the same order.

We should also estimate the probability for a upward fluctuation of the energy, i.e. the overlap with a high energy eigenstate.  Fluctuations add in quadratures, so if we act on $N_{\rm modes} = N_{\rm ext}$ modes the overall width is $T_0/N_{\rm ext}^{1/2}$.  This is much larger than the movement ~(\ref{shift}) of the mean, and so is a more important effect.  The probability of a fluctuation $\Delta E$ is given by the central limit theorem as
\be
\label{eq:var}
e^{- N_{\rm ext} \Delta E^2 /T_0^2} \,.
\ee

By acting with all $\theta_i = \pi$, one can change the entanglements for all of these modes.  A single infalling observer still has a limited time for observation, and so will only see a logarithmic number of excited modes.  In fact, with unitaries of this form it is difficult to act differently on the different localized radial modes, and so to send a message of many bits one would want to use multiple species of field and/or multiple partial waves.

The relatively small power of $N$, $2/17$, comes about as follows.  The black hole is stable when its entropy is greater than that of a gas with the same energy confined to the AdS volume.  Because the geometry is ten-dimensional, the entropy of the gas grows rapidly with the radius of the space available.  A more favorable geometry would couple AdS with a large black hole to a one-dimensional auxiliary system as in~\cite{Rocha:2008fe}; this allows one to access $N^2$ modes.

\section{Measures on Hilbert space}

We now make precise the argument in the last paragraph of \S2.1: For any $\epsilon_1 > 0$ and $\epsilon_2 > 0$, there are choices of $\lambda$, $N$, and $\delta$ such that, for at least a fraction $1-\epsilon_1$ of the normalized states $\bra{\psi_+}$ in ${\cal H}_{E_+}$, there is a normalized excited state $\bra{\psi_{\rm e}}$ in $U_I {\cal H}_{E_0}$ such that $|\langle \psi_+ | \psi_{\rm e} \rangle| \geq 1 - \epsilon_2$.  We remind the reader that by the construction in \S2.1, for any equilibrium state in ${\cal H}_{E_0}$ the state $U_I {\cal H}_{E_0}$ has the specific excitations $I$.  The relevant spaces are shown in Fig.~1.
\begin{figure}[!b]
\begin{center}
\vspace {-5pt}
\includegraphics[width=6.2 in]{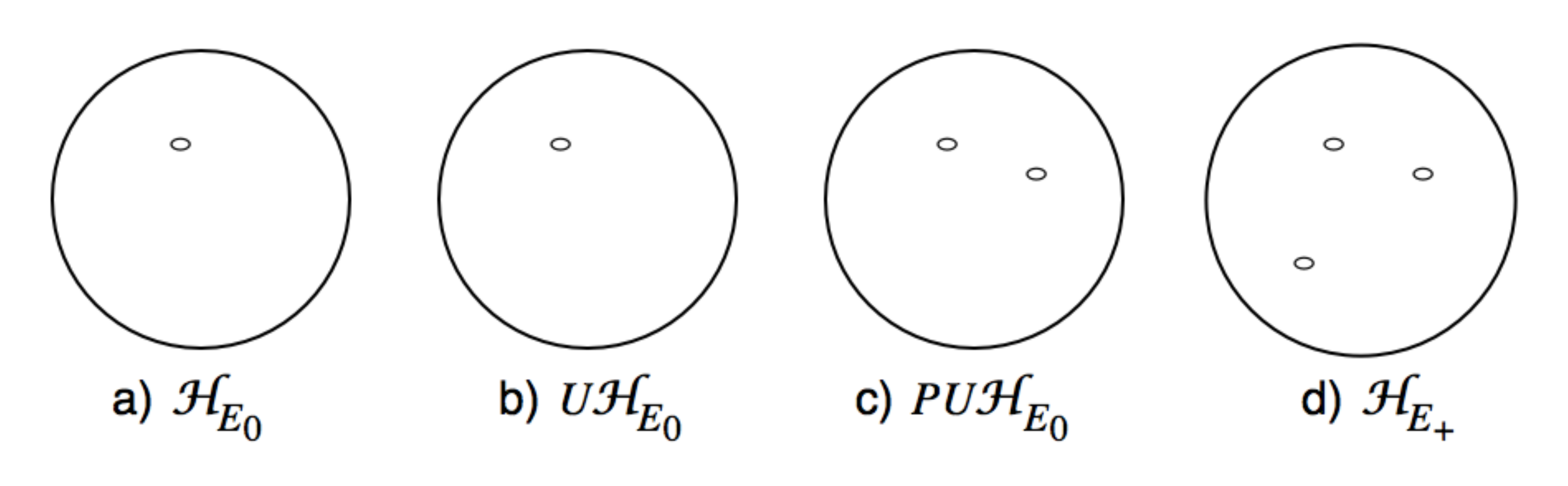}
\end{center}
\vspace {-10pt}
\caption{a) States with energy $E < E_0$.  Atypical states, which are not in the infalling vacuum, are indicated by the small subregion.
b) The image under a unitary $U$.  States outside the subregion are excited. c) The states from (b) projected down to $E < E_+$.  Rare states whose projection is not close to the identity are represented by the second hole.
d) States with energy $E < E_+$, almost all of which are nearly parallel to excited states.  The parametrically rare exceptions are those that project to the two subregions in (c), and those that are not close to the subspace in (c).
}
\label{fig:phase}
\end{figure}

To find $\bra{\psi_{\rm e}}$, we first make the orthogonal decomposition~(\ref{angle}) of ${E_+}$ into $P  U_I  {\cal H}_{E_0}$ and ${\cal H}_\perp = {\cal H}_{E_+}/P  U_I  {\cal H}_{E_0}$,
\be
|\psi_+ \rangle = \cos \theta |\psi_1 \rangle + \sin\theta |\psi_2 \rangle, \quad \cos^2\theta \cong \frac{\dim (P  U_I  {\cal H}_{E_0}) }{\dim ({\cal H}_{E_+} )} = e^{-\delta}\,. \label{angle2}
\ee
A state in $U_I  {\cal H}_{E_0}$ can similarly be decomposed under the orthogonal decomposition into $PU_I  {\cal H}_{E_0}$ and $(1-P)U_I  {\cal H}_{E_0}$.  Defining orthonormal bases $\bra{u_i}$, $\bra{v_j}$, $\bra{w_k}$ for these three spaces, we have
\be
\bra{u_i} = c_{ij} \bra{v_j} + d_{ik} \bra{w_k} \,,\quad \sum_j c^*_{ij} c_{i'j} +  \sum_k d^*_{ik} d_{i'k} = \delta_{ii'} \,.
\ee
By unitary rotations of the $u, v$ bases we can set $c_{ij} = \xi_i \delta_{ij}$ with real $\xi_i \leq 1$, and generically all $\xi_i$ are positive.  We define a generically-norm-preserving map from ${\cal H}_{E_+}$ to $PU_I  {\cal H}_{E_0}$ by mapping $|\psi_+ \rangle$ in (\ref{angle2}) to $ |\psi_1 \rangle$.  We define a map from  $PU_I  {\cal H}_{E_0}$ to $U_I  {\cal H}_{E_0}$ by mapping $\bra{v_i}$ to $\bra{u_i}$.  The composition of the maps defines the candidate $\bra{\psi_{\rm e}}$ associated with $|\psi_+ \rangle $.

The inner product of  $|\psi_+ \rangle $ with the candidate $\bra{\psi_{\rm e}}$   is
\be
\label{eq:IP}
\langle \psi_+| \psi_{\rm e} \rangle = \cos\theta \sum_i \xi_i |\langle v_i |\psi_1  \rangle|^2 \,.
\ee
We are interested in the fraction of states for which \eqref{eq:IP} is greater than $1-\epsilon_2$ for a given $\epsilon_2$; this fraction is at least as great as that for which
\be
\cos\theta > \sqrt{1-\epsilon_2} \, \ \ {\rm and} \ \ ,\quad  \sum_i \xi_i |\langle v_i |\psi_1  \rangle|^2 >  \sqrt{1-\epsilon_2}\,.
\ee
As noted in the main text, $\cos\theta$ is highly peaked at $e^{-\delta/2}$, so that the fraction of states satisfying the first inequality rises to unity very rapidly for $\delta < -\ln( 1-\epsilon_2) \sim \epsilon_2$.  In the second inequality, the mean value of $\xi_i$ will differ from $1$ by an amount that vanishes with increasing $\lambda$.   There will be rare variations upward for some $\xi_i$, but we have seen that the falloff is gaussian in $\lambda$ as in \eqref{eq:var} for moderate $\lambda$ values.\footnote{And, since we choose smooth modes, at least faster than any power law at larger $\lambda$.}  So we can satisfy the second inequality by taking $\lambda$ to be large.

It follows that appropriate choices of $\lambda$, $N$, and $\delta$ wcan make an arbitrarily large fraction of the states in ${\cal H}_{E_+}$ arbitrarily close to states in $U_I {\cal H}_{E_0}$.
Now, not all of the latter are excited, only those that are the images of near-equilibrium states in ${\cal H}_{E_0}$.  We have not given a qualitative description of this discrepancy, but proponents of state-dependent smoothness for horizons would naturally expect the proportion of nonequilibrium states to fall rapidly with $N$:  an infalling observer perceives the state $|\psi_{\rm e}\rangle$ above as containing excitations with energies that can be greater than the Hawking temperature by a power of $N$, so the Boltzmann factor would provide an exponential suppression.  This effect would then
give only a parametrically small reduction in the fraction of states in ${\cal H}_{E_+}$ with excitations comparable to $|\psi_e\rangle$.

\bibliographystyle{JHEP}
\bibliography{statedep}

\end{document}